\RequirePackage{fix-cm}
\documentclass[pdflatex,sn-mathphys-num]{sn-jnl}

\usepackage{graphicx}
\usepackage{amsmath}
\usepackage{amssymb}
\usepackage{xcolor}
\usepackage{manyfoot}
\usepackage{url}
\usepackage{natbib}

\begin{document}

\title[Benchmarking Quantum Annealing for a Greenhouse-Inspired Control QUBO]
{Benchmarking Quantum Annealing for a Greenhouse-Inspired Control QUBO}

\author*[1]{\fnm{Hamzeh} \sur{Alavirad}}\email{h.alavirad@qusmos.com}
\author[1]{\fnm{Maryam Bahrami} \sur{Zanjani}}\email{m.bahrami@qusmos.com}

\affil[1]{\orgname{QUSMOS}, \orgaddress{\city{Karlsruhe}, \country{Germany}}}

\abstract{
We benchmark current annealing-based optimization workflows on a greenhouse-inspired quadratic unconstrained binary optimization problem for binary heater scheduling, where the horizon \(H\) denotes the number of hourly control decisions. For the main one-day instance (\(H=24\)), all solver outputs are decoded back into heater schedules and evaluated in the original greenhouse simulator using the same physical objective and feasibility criterion. Classical simulated annealing and path-integral simulated quantum annealing produce feasible near-optimal solutions in all repetitions, with best objectives close to the exact optimum. In contrast, the tested D-Wave Leap Hybrid BQM workflow is less reliable and does not outperform the classical baselines under 15--60~s requested time limits. Direct D-Wave QPU execution on reduced instances remains feasible in all runs and recovers the exact optimum for \(H=10\) and \(H=12\), but the exact-hit rate drops from \(5/10\) to \(2/10\) and then to \(0/10\) at \(H=14\), with substantially higher variance than the classical baselines. The results do not indicate quantum advantage, but provide a reproducible, physically decoded benchmark that exposes the current strengths and limitations of classical, hybrid, and direct quantum annealing workflows on structured control QUBOs.
}


\maketitle

\section{Introduction}

Greenhouse climate control is an important optimization problem because environmental regulation directly affects crop performance, resource consumption, and operating cost. In practical greenhouse operation, climate management depends on several strongly coupled variables, including temperature, humidity, radiation, and $\mathrm{CO}_2$, together with external disturbances, actuator limitations, and economic constraints. This makes greenhouse climate control a complex decision-making problem in which model quality, optimization strategy, and computational tractability all play a central role \cite{Bersani2020, Morcego2023, Chen2025}.

A broad range of methods has been investigated to optimize greenhouse climate control systems. Model-based approaches, in particular model predictive control (MPC), are widely used because they combine process dynamics, operational constraints, and economic or agronomic objectives within a receding-horizon framework \cite{Bersani2020,Morcego2023}. In parallel, data-driven and learning-based approaches have been explored to improve prediction and decision-making under uncertainty \cite{Ajagekar2023,Mansour2025}. Surrogate-based optimization has also been proposed to determine environmental targets from learned crop-response models. In particular, Lu et. al., \cite{Lu2023} introduced an approach based on support vector regression and a quantum genetic algorithm, which should be understood here as a quantum-inspired classical metaheuristic rather than direct execution on quantum hardware. Related earlier work used support-vector-machine-based models for greenhouse light-environment optimization \cite{Xin2019}.

These studies show that greenhouse control has already moved beyond purely rule-based operation toward optimization-driven and learning-assisted formulations.

From the perspective of quantum computing, this class of problems is relevant because discretized control formulations can lead to large structured combinatorial optimization problems. After suitable simplification, binary encoding, and penalty-based reformulation, such problems can be expressed as quadratic unconstrained binary optimization (QUBO) models \cite{Glover2018,DWaveQUBO}. This is important because QUBO provides a common interface between the application-side control problem and several quantum-related optimization paradigms. In the near term, quantum-inspired optimization methods have shown promising behavior on hard combinatorial problems, suggesting that physics-inspired search strategies may be useful even on classical or specialized non-gate-based hardware \cite{Jiang2023}. In parallel, quantum-native optimization approaches such as the Quantum Approximate Optimization Algorithm (QAOA) provide a longer-term route for solving binary optimization problems on gate-based quantum computers \cite{Farhi2014}. Although large-scale fault-tolerant quantum optimization is not yet available in practice, these developments motivate expressing control-oriented benchmark problems in forms that are compatible with both classical and quantum optimization workflows.

Against this background, the present work does not aim to propose a new greenhouse controller or a new agronomic set-point strategy. Instead, it studies greenhouse-inspired control benchmarks formulated as QUBO problems and evaluates them with exact, classical, and D-Wave solution workflows \cite{Glover2018,DWaveQUBO,DWaveHybrid}. The main benchmark formulation is solved with exact, classical, and D-Wave hybrid methods under a common logical structure, while direct-QPU experiments are performed on reduced instances of the same benchmark class to account for current hardware and embedding limitations \cite{DWaveHybrid,DWaveDecomposing}. This distinction is important because direct execution on D-Wave quantum hardware requires minor embedding of the logical binary quadratic model into the physical qubit graph, and the quality of this mapping can affect observed solution quality \cite{DWaveEmbeddingIntro,DWaveEmbeddingGuidance}. Accordingly, the contribution of this paper is a practical benchmark of structured greenhouse-inspired QUBO formulations on current D-Wave workflows, with emphasis on feasibility, solution quality, and realistic benchmarking rather than claims of quantum advantage \cite{King2015}.

\section{Greenhouse-inspired benchmark model}
\label{sec:greenhouse-toy-model}

We consider a reduced greenhouse-inspired benchmark with a single binary actuator, namely a heater. At time step $k$, the control decision is
\begin{equation}
u_k \in \{0,1\},
\label{eq:toy_control}
\end{equation}
where $u_k=1$ denotes heater on and $u_k=0$ heater off. The system state consists of the air temperature $T_{\mathrm{air},k}$ and a slower body or canopy temperature $T_{\mathrm{body},k}$.

For the benchmark formulation, the state evolution is described in discrete time by
\begin{equation}
\mathbf{x}_{k+1}
=
A_d \mathbf{x}_k
+
B_u u_k
+
B_{\mathrm{out}} T_{\mathrm{out},k}
+
B_{\mathrm{sol}} S_k,
\qquad
\mathbf{x}_k=
\begin{bmatrix}
T_{\mathrm{air},k}\\
T_{\mathrm{body},k}
\end{bmatrix},
\label{eq:toy_dynamics}
\end{equation}
where $A_d$ is the discrete-time state-transition matrix, $B_u$ is the heater input vector, $B_{\mathrm{out}}$ is the disturbance vector associated with outdoor temperature $T_{\mathrm{out},k}$, and $B_{\mathrm{sol}}$ is the disturbance vector associated with solar radiation $S_k$. The initial condition is denoted by $\mathbf{x}_0$.

Plant performance is represented through a reduced growth proxy. First, an effective growth temperature is defined as
\begin{equation}
T_{\mathrm{grow},k}
=
\alpha T_{\mathrm{air},k}
+
(1-\alpha)T_{\mathrm{body},k},
\label{eq:toy_growth_temp}
\end{equation}
where $\alpha \in [0,1]$ is a weighting factor. Second, a bounded light factor is introduced as
\begin{equation}
L_k=
\frac{S_k}{S_k+K_{\mathrm{light}}},
\label{eq:toy_light_factor}
\end{equation}
where $K_{\mathrm{light}}$ is a light-saturation parameter. The instantaneous growth proxy is then written as
\begin{equation}
g_k
=
L_k
\Bigl(
g_{\max}
-
\eta
\bigl(
T_{\mathrm{grow},k}-T^\star
\bigr)^2
\Bigr),
\label{eq:toy_growth}
\end{equation}
where $g_{\max}$ is the maximum growth value, $\eta$ is the curvature parameter, and $T^\star$ is the preferred growth temperature. The growth proxy is not clamped to zero. Therefore, for sufficiently large deviations from $T^\star$, the instantaneous value $g_k$ can become negative. In the reported feasible benchmark solutions, the temperature trajectories remain within the prescribed bounds and do not exploit this negative-growth regime. The proxy should therefore be interpreted as a quadratic performance score rather than a calibrated agronomic growth model.

Energy use and electricity cost are computed diagnostically as
\begin{equation}
E_k=P_{\mathrm{heater}}u_k\Delta t,
\qquad
C_k=\lambda_E p_k E_k,
\label{eq:toy_energy_cost}
\end{equation}
where $P_{\mathrm{heater}}$ is the heater power, $\Delta t$ is the sampling interval, $p_k$ is the electricity tariff, and $\lambda_E$ is an energy-cost scaling factor.

Climate feasibility is assessed on the air-temperature trajectory through
\begin{equation}
T_{\min}\le T_{\mathrm{air},k}\le T_{\max}
\qquad \forall k,
\label{eq:toy_bounds}
\end{equation}
where $T_{\min}$ and $T_{\max}$ denote the admissible lower and upper temperature bounds. The benchmark therefore captures, in reduced form, the trade-off between climate regulation, growth-oriented performance, and energy use, while remaining structured enough for exact, classical, hybrid, and quantum-annealing-based optimization.

Further details of the toy-model construction, including the continuous-time thermal model, exogenous input generation, thermostat baseline, and benchmark parameter values, are given in Appendix~\ref{app:toy-model-details}.

\section{QUBO formulation}
\label{sec:qubo-formulation}

For the optimization problem, the exogenous trajectories $T_{\mathrm{out},k}$, $S_k$, and $p_k$ are treated as known data, and the heater schedule
\begin{equation}
u_k \in \{0,1\}, \qquad k=0,\dots,H-1,
\label{eq:qubo_control}
\end{equation}
is taken as the decision variable. In the benchmark formulation, the no-lag case is assumed, so that
\begin{equation}
u_{\mathrm{applied},k}=u_k,
\label{eq:qubo_applied_control}
\end{equation}
which is consistent with the no-lag case in \eqref{eq:app_toy_applied_nolag}. This no-lag assumption is used to keep the benchmark focused on the combinatorial scheduling problem rather than actuator-response modelling. In real greenhouse systems, heating dynamics and actuator delays may be relevant; these can be incorporated by replacing \eqref{eq:qubo_applied_control} with the lagged mapping described in Appendix~\ref{app:toy-model-details}.

The optimization problem is written directly in QUBO form as
\begin{equation}
Q(\mathbf{z})
=
Q_{\mathrm{growth}}
+
Q_{\mathrm{energy}}
+
Q_{\mathrm{switch}}
+
Q_{\mathrm{bound}},
\label{eq:qubo_total}
\end{equation}
to be minimized over the binary vector
\begin{equation}
\mathbf{z}
=
\left(
u_0,\dots,u_{H-1},
\{z_{k,m}^{L}\},
\{z_{k,m}^{U}\}
\right)^\top
\in \{0,1\}^n.
\label{eq:qubo_binary_vector}
\end{equation}
Here, $Q_{\mathrm{growth}}$ represents the negative growth contribution, so that minimizing it favors temperatures close to the preferred growth regime; $Q_{\mathrm{energy}}$ accounts for electricity consumption; $Q_{\mathrm{switch}}$ penalizes excessive heater switching; and $Q_{\mathrm{bound}}$ enforces the hard air-temperature bounds through quadratic penalties with binary slack variables.

To obtain the individual terms in \eqref{eq:qubo_total}, the discrete-time dynamics \eqref{eq:toy_dynamics} are unfolded over the horizon. Since the model is linear and the exogenous inputs are fixed, the state at step $k$ can be written as an affine function of the binary heater sequence,
\begin{equation}
\mathbf{x}_k
=
\bar{\mathbf{x}}_k
+
\sum_{t=0}^{H-1}\Gamma_{k,t}u_t,
\qquad k=0,\dots,H,
\label{eq:qubo_state_affine}
\end{equation}
where $\bar{\mathbf{x}}_k \in \mathbb{R}^2$ collects the contribution of the initial state and the exogenous inputs, and $\Gamma_{k,t}\in\mathbb{R}^{2\times 1}$ is the state sensitivity with respect to the binary control $u_t$. The air temperature and growth temperature then follow from \eqref{eq:toy_dynamics} and \eqref{eq:toy_growth_temp} as
\begin{equation}
T_{\mathrm{air},k}
=
\mathbf{e}_{\mathrm{air}}^\top \mathbf{x}_k
=
\bar T^{\mathrm{air}}_k
+
\sum_{t=0}^{H-1}\gamma^{\mathrm{air}}_{k,t}u_t,
\label{eq:qubo_air_affine}
\end{equation}
\begin{equation}
T_{\mathrm{grow},k}
=
\mathbf{e}_{\mathrm{grow}}^\top \mathbf{x}_k
=
\bar T^{\mathrm{grow}}_k
+
\sum_{t=0}^{H-1}\gamma^{\mathrm{grow}}_{k,t}u_t,
\label{eq:qubo_grow_affine}
\end{equation}
where $\mathbf{e}_{\mathrm{air}}=[1,\,0]^\top$, $\mathbf{e}_{\mathrm{grow}}=[\alpha,\,1-\alpha]^\top$, $\bar T^{\mathrm{air}}_k=\mathbf{e}_{\mathrm{air}}^\top\bar{\mathbf{x}}_k$, $\bar T^{\mathrm{grow}}_k=\mathbf{e}_{\mathrm{grow}}^\top\bar{\mathbf{x}}_k$, $\gamma^{\mathrm{air}}_{k,t}=\mathbf{e}_{\mathrm{air}}^\top\Gamma_{k,t}$, and $\gamma^{\mathrm{grow}}_{k,t}=\mathbf{e}_{\mathrm{grow}}^\top\Gamma_{k,t}$.

Using \eqref{eq:toy_growth}, \eqref{eq:toy_light_factor}, and \eqref{eq:qubo_grow_affine}, the negative growth contribution in \eqref{eq:qubo_total} can be written, up to an additive constant, as
\begin{equation}
Q_{\mathrm{growth}}
=
\eta \Delta t
\sum_{k=0}^{H-1}
L_k
\left(
\bar T^{\mathrm{grow}}_k
+
\sum_{t=0}^{H-1}\gamma^{\mathrm{grow}}_{k,t}u_t
-
T^\star
\right)^2.
\label{eq:qubo_growth_term}
\end{equation}
The energy term follows directly from \eqref{eq:toy_energy_cost} and \eqref{eq:qubo_applied_control},
\begin{equation}
Q_{\mathrm{energy}}
=
\lambda_E P_{\mathrm{heater}}\Delta t
\sum_{k=0}^{H-1} p_k u_k,
\label{eq:qubo_energy_term}
\end{equation}
and the switching term is written in quadratic binary form as
\begin{equation}
Q_{\mathrm{switch}}
=
\lambda_S
\sum_{k=1}^{H-1}
\left(
u_k+u_{k-1}-2u_ku_{k-1}
\right).
\label{eq:qubo_switch_term}
\end{equation}

For binary variables, the expression satisfies
\(
u_k+u_{k-1}-2u_ku_{k-1}=|u_k-u_{k-1}|,
\)
because it is equal to zero when the two consecutive heater states are identical and equal to one when they differ. Thus, \eqref{eq:qubo_switch_term} penalizes heater switching events while remaining quadratic.

The hard air-temperature bounds in \eqref{eq:toy_bounds} are incorporated through quadratic penalties with binary slack variables. For each $k=1,\dots,H$, we impose
\begin{equation}
T_{\mathrm{air},k}-T_{\min}-s_k^{L}=0,
\qquad
T_{\max}-T_{\mathrm{air},k}-s_k^{U}=0,
\label{eq:qubo_slack_constraints}
\end{equation}
with nonnegative slack variables represented in binary form as
\begin{equation}
s_k^{L}=\sum_{m=0}^{M-1}2^m\Delta_s z_{k,m}^{L},
\qquad
s_k^{U}=\sum_{m=0}^{M-1}2^m\Delta_s z_{k,m}^{U},
\label{eq:qubo_slack_binary}
\end{equation}
where $z_{k,m}^{L},z_{k,m}^{U}\in\{0,1\}$, $\Delta_s$ is the slack resolution, and $M$ is the number of slack bits. 

The two slack variables encode the lower and upper inequalities separately. At any binary assignment where both penalty residuals in \eqref{eq:qubo_slack_constraints} vanish, the relation
\(s_k^L+s_k^U=T_{\max}-T_{\min}\) holds by construction as a consequence of the shared temperature variable. However, this relation is not imposed as an additional constraint. For approximate solvers, one or both residuals may be nonzero simultaneously; in that case, the decoded temperature is penalized but not explicitly corrected by the slack variables. The formulation should therefore be interpreted as a penalty-based encoding of the two one-sided inequalities. It doubles the slack-bit footprint and may introduce additional coupling structure in the BQM, which is one reason why more compact inequality encodings are identified as future work.

Substituting \eqref{eq:qubo_air_affine} into \eqref{eq:qubo_slack_constraints} yields
\begin{align}
Q_{\mathrm{bound}}
=\;&
A_{\mathrm{bound}}
\sum_{k=1}^{H}
\left(
\bar T^{\mathrm{air}}_k
+
\sum_{t=0}^{H-1}\gamma^{\mathrm{air}}_{k,t}u_t
-
T_{\min}
-
\sum_{m=0}^{M-1}2^m\Delta_s z_{k,m}^{L}
\right)^2
\nonumber\\
&+
A_{\mathrm{bound}}
\sum_{k=1}^{H}
\left(
T_{\max}
-
\bar T^{\mathrm{air}}_k
-
\sum_{t=0}^{H-1}\gamma^{\mathrm{air}}_{k,t}u_t
-
\sum_{m=0}^{M-1}2^m\Delta_s z_{k,m}^{U}
\right)^2,
\label{eq:qubo_bound_term}
\end{align}
where $A_{\mathrm{bound}}>0$ is the penalty weight.

In the numerical benchmarks, we use $M=6$ slack bits per bound and time step, with slack resolution $\Delta_s=0.25^\circ\mathrm{C}$. This gives a maximum representable slack value of
\[
s_{\max}=(2^M-1)\Delta_s = 15.75^\circ\mathrm{C}.
\]
The bound-penalty weight is set to $A_{\mathrm{bound}}=120$ for both lower- and upper-bound constraints. These values were chosen to make temperature-bound violations energetically unfavorable while keeping the slack resolution sufficiently fine for the decoded greenhouse trajectories.

We use standard binary slack encoding for transparency and reproducibility. More compact or unbalanced slack encodings may reduce the number of auxiliary variables and improve direct-QPU embedding efficiency; these alternatives are left for future work.

Equivalently, the problem can be written in standard matrix form as
\begin{equation}
\min_{\mathbf{z}\in\{0,1\}^n}\;
\mathbf{z}^\top Q \mathbf{z} + c,
\label{eq:qubo_matrix_form}
\end{equation}
where $Q$ is the symmetric QUBO matrix and $c$ is a constant offset. This representation is directly compatible with classical QUBO solvers and quantum annealing hardware \cite{Glover2018,DWaveQUBO,DWaveReformulating}.

\section{Solvers and benchmark protocol}
\label{sec:solvers}

We consider four main solution workflows for the $H=24$ benchmark instance, together with additional direct-QPU experiments on reduced instances. As a reference baseline, the decoded control problem is solved exactly by exhaustive enumeration of all binary heater schedules $u \in \{0,1\}^H$. For each candidate schedule, the greenhouse state trajectory is recomputed using the simulator dynamics, and the corresponding objective and feasibility quantities are evaluated directly. This exact reference is computationally restricted to moderate horizons, but it provides the ground-truth optimum for the $H=24$ benchmark used in this study.

The QUBO-based methods solve the binary quadratic model defined in \eqref{eq:qubo_matrix_form}. As a classical baseline, we use \texttt{SimulatedAnnealingSampler}, which performs thermal annealing on the binary quadratic model. As a simulated quantum annealing baseline, we use \texttt{PathIntegralAnnealingSampler}, which implements path-integral simulated quantum annealing on a classical computer.\footnote{We use the term simulated quantum annealing to emphasize that the method is executed classically. Its relationship to quantum-inspired optimization and its comparative advantage over classical simulated annealing are problem-dependent and remain subject to benchmarking.} In addition, we use D-Wave's Leap Hybrid BQM workflow, which operates on the same logical BQM while handling decomposition and solver orchestration internally \cite{DWaveSamplers,DWaveHybrid}. Direct-QPU experiments are performed separately on reduced instances of the same benchmark class. For direct QPU execution, the logical binary quadratic model must be minor-embedded into the QPU connectivity graph, introducing chains and embedding-dependent overhead \cite{DWaveEmbeddingIntro,DWaveSolverParameters,DWaveEmbeddingGuidance,King2015}.

The main benchmark instance uses $H=24$ and $\Delta t=1$~h, corresponding to one day of hourly binary heater scheduling. The initial conditions are $T_{\mathrm{air},0}=18.0^\circ\mathrm{C}$ and $T_{\mathrm{body},0}=17.0^\circ\mathrm{C}$, with a start time of 06:00. With the above slack encoding, this instance yields a BQM with 312 binary variables, consisting of 24 heater variables and 288 slack bits, and 4596 quadratic couplers. For this instance, exact enumeration, classical simulated annealing, path-integral simulated quantum annealing, and the Leap Hybrid BQM workflow are compared under the same decoded objective and feasibility evaluation. The stochastic solvers are executed repeatedly, and the reported values distinguish between best feasible decoded solutions and averages over feasible decoded runs. For the hybrid workflow, a 30~s requested solver time is used as the representative setting in the main comparison. Additional 15~s and 60~s requested time-limit runs are used only as diagnostic checks to test whether the decoded hybrid performance improves monotonically with requested solver time.

\begin{table}[t]
\centering
\caption{Solver and sampler settings used in the benchmark. SA, PIA, hybrid, and direct-QPU workflows are executed using ten independent repetitions. The hybrid settings denote requested Leap Hybrid BQM time limits rather than full end-to-end wall-clock runtimes.}
\label{tab:solver_settings}
\small
\setlength{\tabcolsep}{5pt}
\begin{tabular}{p{0.22\textwidth} p{0.70\textwidth}}
\hline
Workflow & Settings \\
\hline

Exact enumeration
& Exhaustive enumeration of all $2^H$ binary heater schedules, followed by simulator-side decoding and objective evaluation. \\

SA
& \texttt{SimulatedAnnealingSampler}; 10 seeds; \texttt{num\_reads}=5000; \texttt{num\_sweeps}=3000. \\

PIA
& \texttt{PathIntegralAnnealingSampler}; 10 seeds; \texttt{num\_reads}=5000; \texttt{beta\_schedule\_type=custom}; \texttt{Hd\_field=[10,0]}; \texttt{Hp\_field=[0,10]}. \\

Leap Hybrid BQM
& \texttt{hybrid\_binary\_quadratic\_model\_version2p}; 10 independent submissions; requested time limits of 15~s, 30~s, and 60~s. The 30~s setting is used as the representative hybrid result in the main comparison. \\

Direct QPU
& \texttt{EmbeddingComposite(DWaveSampler)}; solver \texttt{Advantage2\_system1}; 10 independent submissions; \texttt{num\_reads}=5000; annealing time 20~$\mu$s; \texttt{auto\_scale=True}. \\

\hline
\end{tabular}
\end{table}

The D-Wave Ocean components used in the reported runs were:
\begin{itemize}
    \item \texttt{dwave-system} v1.34.0,
    \item \texttt{dimod} v0.12.21,
    \item \texttt{dwave-samplers} v1.7.0.
\end{itemize}
The \texttt{dwave-ocean-sdk} meta-package was not installed as a separate package in the execution environment.

All stochastic workflows are evaluated using ten independent repetitions. This number was chosen as a compromise between repeated-run variability and the practical cost of cloud-based hybrid and QPU access. The resulting statistics are therefore intended as benchmark diagnostics rather than high-confidence estimates of the full solver-output distributions. For this reason, the tables report standard deviations where appropriate, and the text avoids statistical claims about time-limit monotonicity or solver superiority beyond the tested settings.

Direct-QPU experiments are performed for reduced horizons $H=10$, $H=12$, and $H=14$. These instances are small enough for direct embedding and execution on the quantum annealer. The purpose of these reduced direct-QPU experiments is not to replace the main $H=24$ benchmark, but to assess the behavior of direct quantum annealing on smaller greenhouse-inspired QUBOs under the same decoding and feasibility-evaluation procedure.

For all QUBO-based methods, the optimization variables are the full binary vector in \eqref{eq:qubo_binary_vector}, including both the heater variables and the slack bits introduced in \eqref{eq:qubo_slack_binary}. Solver comparison, however, is carried out at the level of the decoded heater schedule. After sampling, the control sequence $u$ is extracted from the returned binary vector, the corresponding state trajectory is recomputed using \eqref{eq:toy_dynamics}, and the decoded solution is re-evaluated using the original simulator quantities defined in \eqref{eq:toy_growth}, \eqref{eq:toy_energy_cost}, and \eqref{eq:toy_bounds}. This common post-processing step ensures that all QUBO-based methods are assessed against the same physical model and feasibility criterion rather than only raw BQM energy.

The reported key performance indicators are the decoded objective value, total energy consumption, total growth, air- and body-temperature trajectories, the number of temperature-bound violations, the number of heater switching events, and solver runtime. Total energy consumption is computed as
\begin{equation}
E_{\mathrm{tot}} = \sum_{k=0}^{H-1} E_k,
\label{eq:benchmark_energy_total}
\end{equation}
where $E_k$ is defined in \eqref{eq:toy_energy_cost}. Total growth is computed as
\begin{equation}
G_{\mathrm{tot}} = \sum_{k=0}^{H-1} g_k \Delta t,
\label{eq:benchmark_growth_total}
\end{equation}
where $g_k$ is defined in \eqref{eq:toy_growth}. The number of temperature-bound violations is defined by
\begin{equation}
N_{\mathrm{viol}}=
\sum_{k=0}^{H}
\mathbf{1}\!\left[
T_{\mathrm{air},k}<T_{\min}
\;\vee\;
T_{\mathrm{air},k}>T_{\max}
\right],
\label{eq:benchmark_violations}
\end{equation}
and the number of heater switching events is
\begin{equation}
N_{\mathrm{switch}}=
\sum_{k=1}^{H-1}|u_k-u_{k-1}|.
\label{eq:benchmark_switches}
\end{equation}

The decoded simulator-side objective reported in the results is
\begin{equation}
J_{\mathrm{total}}(u)
=
G_{\mathrm{tot}}(u)
-
J_E(u)
-
J_S(u),
\label{eq:benchmark_jtotal}
\end{equation}
with
\begin{equation}
J_E(u)
=
\lambda_E
\sum_{k=0}^{H-1} p_k E_k,
\qquad
J_S(u)
=
\lambda_S N_{\mathrm{switch}}.
\label{eq:benchmark_decoded_cost_terms}
\end{equation}

This decoded objective differs from the raw QUBO energy $Q(\mathbf{z})$. In particular, the bound-penalty term $Q_{\mathrm{bound}}$ is not part of the physical performance objective; instead, temperature-bound satisfaction is evaluated separately through $N_{\mathrm{viol}}$. Therefore, all solvers are compared using the same decoded physical objective and feasibility criterion rather than raw BQM energy.

For instances with a known exact optimum, objective gaps relative to the exact decoded solution are additionally evaluated. For D-Wave hybrid runs, the reported timing refers to the requested hybrid solver time limit and the corresponding D-Wave reported solver time, excluding external queueing, network latency, and notebook overhead. The 15~s, 30~s, and 60~s settings should therefore be interpreted as requested hybrid time limits rather than complete end-to-end wall-clock runtimes. For direct-QPU runs, the reported hardware timing includes the QPU access time returned by the D-Wave system. Because direct-QPU performance depends on embedding and sampling details, these results are reported separately from the main $H=24$ benchmark.

The purpose of this benchmark is not to establish quantum advantage, but to assess how exact, classical, quantum-inspired, hybrid, and direct quantum-annealing workflows behave on the same structured control benchmark under a common decoding and evaluation procedure.

\section{Results}
\label{sec:results}

This section reports the benchmark results for the greenhouse-inspired QUBO instances. The main benchmark uses a horizon of $H=24$ with sampling interval $\Delta t=1$~h. This instance is solved using exact enumeration, classical simulated annealing (SA), path-integral simulated quantum annealing (PIA), and D-Wave's Leap Hybrid BQM workflow. Direct-QPU experiments are reported separately for reduced instances with $H=10$, $H=12$, and $H=14$, where direct embedding on the quantum annealer remains practical.

The objective value $J_{\mathrm{total}}$ reported in this section is the decoded simulator-side performance measure, where larger values indicate better solutions. Feasibility is assessed through the air-temperature bound constraint, and a decoded solution is counted as feasible only if the number of temperature-bound violations is zero. For stochastic solvers, both the best feasible run and the mean over feasible runs are reported.

\subsection*{Main $H=24$ benchmark}
\label{subsec:results_main}

Table~\ref{tab:main_kpi_comparison} summarizes the main benchmark comparison for the $H=24$ instance. Exact enumeration provides the reference optimum for this instance. Classical SA and PIA both return feasible decoded solutions in all ten repetitions. The D-Wave Leap Hybrid BQM solver with a 30~s time limit returns feasible decoded solutions in seven out of ten repetitions.

\begin{table}[t]
\centering
\caption{Main benchmark comparison for the $H=24$, $\Delta t=1$~h instance. The benchmark uses $T_{\mathrm{air},0}=18.0^\circ\mathrm{C}$, $T_{\mathrm{body},0}=17.0^\circ\mathrm{C}$, $T_{\min}=16^\circ\mathrm{C}$, $T_{\max}=26^\circ\mathrm{C}$, $T^\star=22^\circ\mathrm{C}$, $\lambda_E=1.5$, and $\lambda_S=0.5$. For stochastic solvers, Best $J$, $E$ [kWh], Growth, Switch, and Runtime refer to the best feasible decoded run. The mean objective and standard deviation are computed over feasible stochastic runs. The objective $J_{\mathrm{total}}$ is maximized. Exact enumeration is deterministic.}
\label{tab:main_kpi_comparison}
\small
\setlength{\tabcolsep}{4pt}
\begin{tabular}{lccccccc}
\hline
Solver
& Feas.
& \begin{tabular}[c]{@{}c@{}}Best\\$J$\end{tabular}
& \begin{tabular}[c]{@{}c@{}}Mean $J$\\$\pm\,\sigma_J$\end{tabular}
& \begin{tabular}[c]{@{}c@{}}$E$\\{[kWh]}\end{tabular}
& Growth
& Switch
& \begin{tabular}[c]{@{}c@{}}Runtime\\{[s]}\end{tabular} \\
\hline

Exact
& N/A
& -145.14
& $-145.14$
& 1100
& 40.36
& 11
& 1028.27 \\

SA
& 10/10
& -150.11
& $-161.60 \pm 6.76$
& 1100
& 35.39
& 11
& 31.84 \\

PIA
& 10/10
& -149.11
& $-169.94 \pm 8.45$
& 1100
& 35.39
& 9
& 3.60 \\

\begin{tabular}[c]{@{}l@{}}Leap Hybrid\\BQM, 30~s\end{tabular}
& 7/10
& -174.32
& $-203.84 \pm 16.72$
& 1300
& 41.68
& 12
& 29.99 \\

\hline
\end{tabular}
\end{table}

Exact enumeration provides the reference optimum by construction, with $J_{\mathrm{total}}=-145.14$ for the $H=24$ benchmark. Classical SA reaches a best objective of $-150.11$, corresponding to an objective gap of approximately $4.97$ relative to the exact solution. PIA performs similarly in its best run, reaching $-149.11$, with an objective gap of approximately $4.03$. Both SA and PIA therefore provide feasible near-optimal solutions for this benchmark instance.

The Leap Hybrid BQM result is weaker under the tested configuration. Its best feasible decoded solution reaches $J_{\mathrm{total}}=-174.32$, corresponding to a gap of approximately $29.18$ relative to the exact optimum. The mean feasible hybrid objective, $-203.84$, is also substantially below the corresponding SA and PIA means. Therefore, for this $H=24$ instance, the tested hybrid workflow does not outperform the classical annealing baselines.

The hybrid result is nevertheless informative. The feasible hybrid runs satisfy the temperature constraints and have zero violations. The performance loss is therefore not primarily a feasibility failure, but rather a weaker trade-off between growth, energy use, and switching. In particular, the best hybrid solution achieves a larger growth proxy than the exact solution, $41.68$ compared with $40.36$, but does so by using more energy, $1300$~kWh compared with $1100$~kWh, and by producing a worse overall objective. This indicates that the decoded hybrid solution remains physically meaningful but is not optimal under the chosen objective weighting.

Since $\Delta t=1$~h and $P_{\mathrm{heater}}=100$~kW, the total energy value can be cross-checked directly from the number of heater-on intervals: each heater-on step contributes 100~kWh. Thus, the exact solution energy of 1100~kWh corresponds to 11 heater-on intervals, while the best hybrid solution energy of 1300~kWh corresponds to 13 heater-on intervals.

The penalty weight $A_{\mathrm{bound}}$ was selected empirically to prioritize feasibility of the decoded temperature trajectory. This choice makes temperature-bound violations energetically unfavorable, but it can also reshape the BQM energy landscape and affect the relative balance between growth, energy, and switching terms. The weaker hybrid performance should therefore not be interpreted as an intrinsic limitation of hybrid quantum-classical optimization. It may also reflect the present penalty scaling, the dense coupling structure induced by the rollout dynamics and slack variables, and the mismatch between internal BQM energy minimization and the decoded simulator-side objective.

\subsection*{Hybrid time-limit diagnostic}
\label{subsec:results_hybrid_time_limit}

To test whether the weaker hybrid result was caused by insufficient requested solver time, the $H=24$ hybrid experiment was evaluated using requested time limits of 15~s, 30~s, and 60~s. Table~\ref{tab:hybrid_time_limit_comparison} compares the 15~s, 30~s and 60~s hybrid settings.

\begin{table}[t]
\centering
\caption{Diagnostic comparison of Leap Hybrid BQM results for the $H=24$ instance using different requested hybrid time limits. Only feasible decoded runs are included in the best and mean objective values. The ``No feasible'' column counts submissions for which no feasible decoded heater schedule was retained under the post-processing criterion.}
\label{tab:hybrid_time_limit_comparison}
\small
\setlength{\tabcolsep}{4pt}
\begin{tabular}{lcccccc}
\hline
Hybrid setting
& Feasible
& No feasible
& \begin{tabular}[c]{@{}c@{}}Best\\$J_{\mathrm{total}}$\end{tabular}
& \begin{tabular}[c]{@{}c@{}}Mean\\$J_{\mathrm{total}}$\end{tabular}
& \begin{tabular}[c]{@{}c@{}}Best\\$E_{\mathrm{tot}}$\\{[kWh]}\end{tabular}
& \begin{tabular}[c]{@{}c@{}}Mean\\$E_{\mathrm{tot}}$\\{[kWh]}\end{tabular} \\
\hline
15~s, 10 repeats
& 5/10
& 5/10
& -181.88
& -194.35
& 1300
& 1360 \\

30~s, 10 repeats
& 7/10
& 3/10
& -174.32
& -203.84
& 1300
& 1400 \\

60~s, 10 repeats
& 2/10
& 8/10
& -185.32
& -200.77
& 1400
& 1500 \\
\hline
\end{tabular}
\end{table}

The hybrid time-limit diagnostic should be interpreted cautiously because each setting was evaluated using only ten independent hybrid submissions. The 30~s setting gives the highest feasible-run rate and the best hybrid objective among the tested settings. However, the non-monotonic behavior from 15~s to 30~s to 60~s should not be interpreted as evidence that longer hybrid runs are generally worse. Rather, it shows that, for this formulation and sample size, increasing the requested hybrid time limit did not produce a reliable improvement in decoded simulator-side performance.

The infeasible hybrid submissions in Table~\ref{tab:hybrid_time_limit_comparison} correspond to submissions for which no feasible decoded heater schedule was retained under the post-processing criterion. Therefore, the diagnostic reflects both the internal BQM search and the external decoded-feasibility filter. A larger number of independent hybrid submissions would be required to distinguish stochastic sampling effects from systematic time-limit behavior.

For this reason, the 30~s setting is retained as the representative hybrid result in the main benchmark comparison, while the 15~s and 60~s settings are treated as diagnostic time-limit checks.

\subsection*{Temperature trajectories}
\label{subsec:results_temperature}

\begin{figure}[ht]
    \centering
    \includegraphics[width=\textwidth]{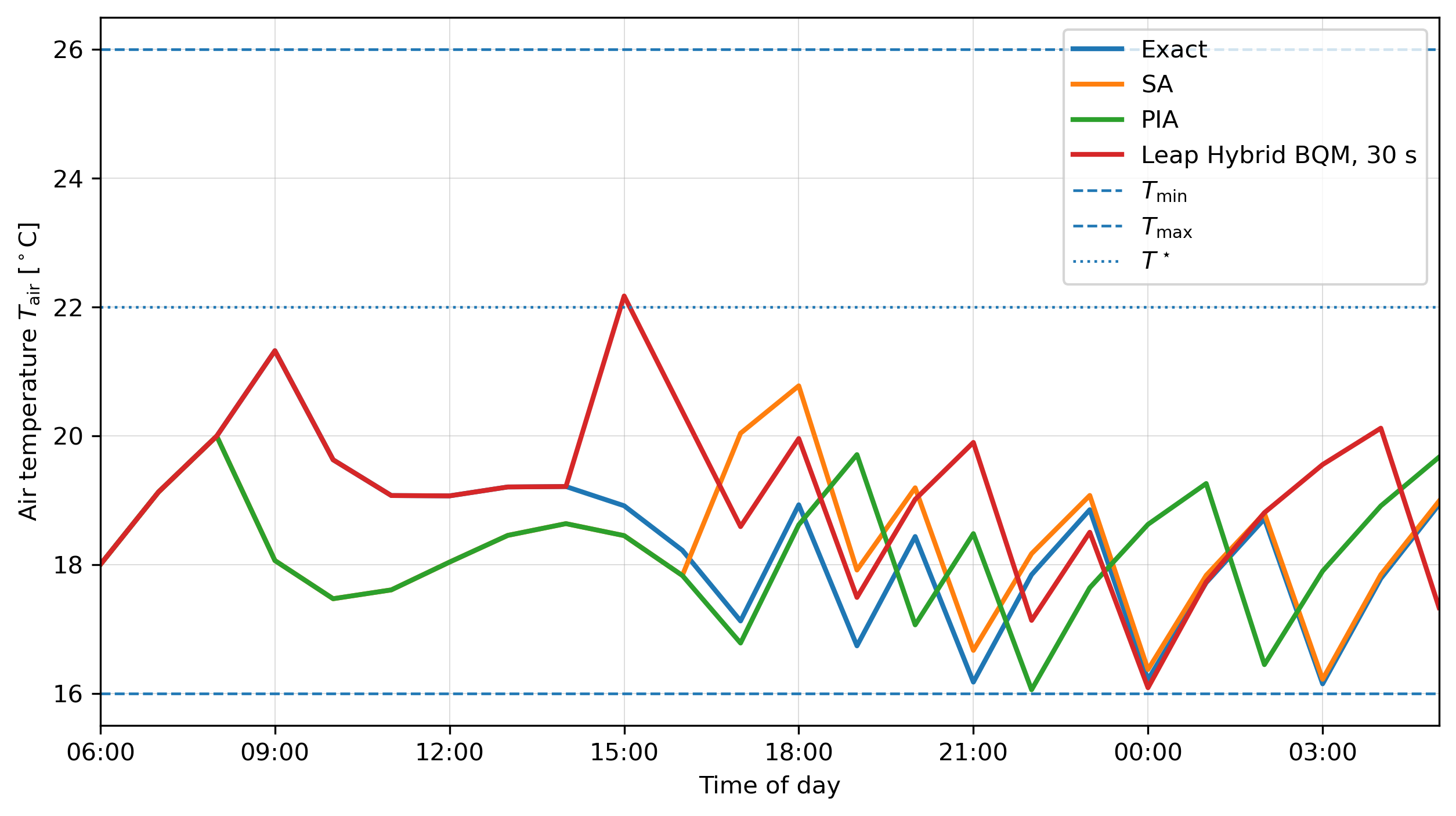}
    \caption{Best feasible air-temperature trajectories for the $H=24$ benchmark. Only the best feasible decoded solution of each solver workflow is shown to keep the comparison readable. The admissible interval is defined by $T_{\min}=16^\circ\mathrm{C}$ and $T_{\max}=26^\circ\mathrm{C}$.}
    \label{fig:air_temperature_comparison}
\end{figure}

Figure~\ref{fig:air_temperature_comparison} compares the air-temperature trajectories for the best feasible decoded solutions of the benchmark solvers. All reported feasible solutions remain within the admissible temperature interval $[T_{\min},T_{\max}]$, consistent with the zero-violation counts in Table~\ref{tab:main_kpi_comparison}. The exact, SA, and PIA solutions follow broadly similar feasible trajectories, whereas the hybrid solution shows a different heater-timing pattern that leads to higher energy use and a lower objective value.

The corresponding body-temperature trajectories and cumulative-growth curves are provided in Appendix~\ref{app:additional_trajectory_diagnostics}. The body-temperature trajectories are smoother and less discriminative than the air-temperature trajectories because of the slower thermal response of the second state. Similarly, the cumulative-growth curves mainly support the quantitative growth--energy trade-off already reported in Table~\ref{tab:main_kpi_comparison}. They are therefore included as supplementary diagnostics rather than central results.

\subsection*{Reduced-instance direct-QPU results}
\label{subsec:results_qpu}

Direct-QPU experiments were performed on reduced instances with $H=10$, $H=12$, and $H=14$. These cases are small enough to allow direct execution on the D-Wave quantum annealer after embedding. The purpose of these experiments is not to solve the largest benchmark instance, but to assess how direct QPU execution behaves on reduced greenhouse-inspired QUBOs under the same decoding and feasibility-evaluation procedure.

Table~\ref{tab:qpu_scaling_results} summarizes the direct-QPU results. For each horizon, the exact objective is reported together with the best and mean decoded QPU objective over ten repetitions. The exact-hit count denotes the number of QPU repetitions that recovered the exact decoded optimum.

\begin{table}[t]
\centering
\small
\caption{Reduced-instance benchmark results for $H=10$, $H=12$, and $H=14$. The table compares exact enumeration, SA, PIA, and direct QPU execution on the same reduced instances. Stochastic means and standard deviations are computed over feasible decoded runs. All stochastic runs were feasible. QPU exact hits and QPU access time are reported only for direct-QPU rows.}
\label{tab:qpu_scaling_results}
\begin{tabular}{c l r r r r c c}
\hline
$H$ 
& Method 
& Vars. 
& Coupl. 
& Best $J$ 
& Mean $J \pm \sigma_J$ 
& QPU hits 
& QPU access [s] \\
\hline

10 & Exact 
& 130 & 1005 
& 8.19 
& -- 
& -- 
& -- \\

10 & SA 
& 130 & 1005 
& 8.19 
& $8.19 \pm 0.00$ 
& -- 
& -- \\

10 & PIA 
& 130 & 1005 
& 8.19 
& $8.19 \pm 0.00$ 
& -- 
& -- \\

10 & QPU 
& 130 & 1005 
& 8.19 
& $-1.04 \pm 12.71$ 
& 5/10 
& 0.85 \\

\hline

12 & Exact 
& 156 & 1362 
& -6.61 
& -- 
& -- 
& -- \\

12 & SA 
& 156 & 1362 
& -6.61 
& $-6.61 \pm 0.00$ 
& -- 
& -- \\

12 & PIA 
& 156 & 1362 
& -6.61 
& $-6.79 \pm 0.38$ 
& -- 
& -- \\

12 & QPU 
& 156 & 1362 
& -6.61 
& $-20.53 \pm 18.07$ 
& 2/10 
& 0.88 \\

\hline

14 & Exact 
& 182 & 1771 
& -34.61 
& -- 
& -- 
& -- \\

14 & SA 
& 182 & 1771 
& -34.61 
& $-35.19 \pm 0.49$ 
& -- 
& -- \\

14 & PIA 
& 182 & 1771 
& -35.11 
& $-36.74 \pm 2.33$ 
& -- 
& -- \\

14 & QPU 
& 182 & 1771 
& -35.11 
& $-58.83 \pm 23.38$ 
& 0/10 
& 0.89 \\

\hline
\end{tabular}
\end{table}

For $H=10$, SA, PIA, and the direct QPU all recover the exact optimum in their best runs. However, the mean QPU objective is substantially below the exact value, indicating a broad distribution of decoded QPU solution quality despite the 5/10 exact-hit rate. For $H=12$, the QPU still recovers the exact optimum in two out of ten repetitions, while SA and PIA remain close to the exact solution in both best and mean performance. For $H=14$, no QPU exact hit is observed. The best QPU solution has a small objective gap of approximately $0.50$, but the mean QPU gap is approximately $24.22$, indicating low reliability at this instance size. The contrast between the near-optimal best QPU sample and the much lower mean QPU objective indicates a highly dispersed decoded-solution distribution with high variance, especially at $H=14$, whereas PIA remains comparatively concentrated around near-optimal decoded objectives.

The reduced-instance results therefore show that direct QPU execution can produce feasible decoded solutions for all tested horizons, but its reliability decreases with horizon and remains weaker than the classical annealing baselines in mean decoded objective. The degradation with increasing horizon appears mainly as a loss of optimality reliability rather than a loss of feasibility.

\subsection*{Discussion of solver behavior}
\label{subsec:results_solver_behavior}

The combined results lead to three observations. First, the greenhouse-inspired QUBO benchmark produces physically interpretable decoded solutions across all solver workflows. Feasibility can be evaluated directly in the simulator, and all feasible decoded solutions satisfy the imposed temperature bounds.

Second, for the exactly benchmarkable $H=24$ instance, the classical annealing baselines are strong. Both SA and PIA produce feasible near-optimal solutions in all repetitions and outperform the tested Leap Hybrid BQM configuration. The hybrid workflow should therefore not be presented as superior for this benchmark instance. Its value in the present study is instead diagnostic: it shows that D-Wave's scalable hybrid BQM workflow can be applied to the structured greenhouse QUBO, but that its performance is sensitive to the formulation, scaling, and solver interaction.

A limitation of this comparison is that $H=24$ is still a moderate benchmark size and is not the regime for which large-scale hybrid decomposition workflows are primarily designed. The choice of $H=24$ was deliberate because it allows comparison against a known exact decoded optimum. However, this also means that the present results should not be interpreted as a definitive assessment of the Leap Hybrid BQM solver on larger multi-day greenhouse horizons. Larger instances, such as $H=72$ or $H=168$, should be studied in future work using strong classical heuristics as baselines, since exact enumeration is no longer practical in that regime.

Third, the direct-QPU results provide a complementary hardware-facing benchmark. Direct QPU execution is feasible for reduced instances and can recover exact optima for $H=10$ and $H=12$. At $H=14$, the QPU no longer recovers the exact optimum in the tested repetitions, although the best decoded solution remains close to optimal. This supports a cautious interpretation: for the tested reduced instances, direct QPU execution produced feasible decoded solutions and occasionally recovered the exact optimum, but its mean decoded performance remained weaker than the classical annealing baselines. The tested instances therefore do not support a claim of quantum advantage.

The hybrid time-limit diagnostic is also limited by the number of independent submissions. Each time-limit setting was evaluated with ten repetitions, which is sufficient to reveal non-monotonic behavior in this particular experiment but not sufficient to determine whether the effect is systematic. The 15~s, 30~s, and 60~s comparison should therefore be interpreted as a diagnostic observation rather than a statistically robust statement about hybrid runtime scaling.

Overall, the results support the role of this greenhouse-inspired QUBO as a structured benchmark for annealing-based optimization workflows. They also show the importance of comparing quantum and quantum-hybrid workflows against strong classical baselines under the same decoded objective and feasibility criteria.

\section{Limitations}
\label{sec:limitations}

The benchmark is intentionally simplified. It contains a single binary heater actuator and does not model ventilation, humidity, $\mathrm{CO}_2$, crop-specific physiology, actuator saturation, forecast uncertainty, or real greenhouse data. The growth term is a quadratic proxy and should not be interpreted as a calibrated biological growth model.

The main hybrid comparison is restricted to $H=24$, where exact enumeration is still possible. This choice enables comparison against a known decoded optimum, but it does not probe the large-scale regime for which hybrid decomposition workflows are primarily designed. Larger multi-day horizons should therefore be studied in future work using strong classical heuristics and mathematical-optimization baselines.

The QUBO formulation uses penalty-based bound enforcement with binary slack variables. The penalty value $A_{\mathrm{bound}}=120$ was selected empirically to prioritize feasible decoded trajectories, but no systematic penalty-sensitivity study is included. Different penalty scalings or alternative inequality encodings may change solver behavior, especially for hybrid and direct-QPU workflows.

The direct-QPU experiments are restricted to reduced horizons $H=10$, $H=12$, and $H=14$ because direct hardware execution requires minor embedding of the logical BQM. These experiments should therefore be interpreted as reduced embedding-aware hardware tests, not as evidence of practical quantum advantage.

The direct-QPU experiments use the default \texttt{EmbeddingComposite} behavior and do not manually tune the chain strength. Chain-break fractions, physical-qubit counts after embedding, and embedding-quality statistics were not recorded systematically for the reported runs. Consequently, the present results cannot separate degradation due to increasing logical problem difficulty from degradation due to embedding quality or chain-breaking effects. This is a limitation of the hardware-facing analysis and should be addressed in future direct-QPU benchmarks.

Finally, the benchmark uses a coarse sampling interval of $\Delta t=1$~h. This is adequate for a structured proof-of-concept benchmark but not for operational greenhouse climate control, where shorter control intervals and actuator-response dynamics would be required.

\section{Conclusion}
\label{sec:conclusion}

This paper introduced a greenhouse-inspired control benchmark formulated as a quadratic unconstrained binary optimization problem and evaluated it using exact enumeration, classical simulated annealing, path-integral simulated quantum annealing, D-Wave's Leap Hybrid BQM solver, and direct QPU execution on reduced instances. The aim was not to propose a deployable greenhouse controller or to claim quantum advantage, but to construct an application-motivated benchmark that exposes the practical behavior of current annealing-based optimization workflows on a structured control problem.

The results show that the benchmark is useful for separating feasibility, decoded solution quality, and hardware scalability. For the $H=24$ instance, exact enumeration provides the reference optimum. Classical SA and PIA produce feasible near-optimal solutions in all repetitions, with best objectives close to the exact solution. Under the tested settings, the Leap Hybrid BQM workflow does not outperform these classical baselines. In the limited ten-repetition diagnostic, increasing the requested hybrid time limit from 30~s to 60~s did not improve the decoded solution quality. This suggests that the limitation is not merely insufficient runtime, but may be related to QUBO scaling, penalty tuning, formulation density, or solver-specific behavior.

The reduced direct-QPU experiments provide a complementary hardware-level result. Direct QPU execution recovers the exact optimum for $H=10$ and $H=12$, with exact-hit rates of $5/10$ and $2/10$, respectively. At $H=14$, no exact hit is observed under the tested settings, although the best decoded QPU solution remains close to the exact optimum. All decoded QPU solutions are feasible and have zero temperature-bound violations. Thus, the main degradation with increasing horizon is a reduction in optimality reliability rather than a loss of feasibility.

Taken together, the results support a cautious interpretation of current quantum annealing workflows for greenhouse-inspired control QUBOs. Direct QPU execution provides a reduced embedding-aware hardware test, while classical annealing methods remain stronger baselines for the tested instance sizes. Hybrid quantum-classical workflows remain relevant for larger or more complex instances, but the present results show that their performance cannot be assumed and must be benchmarked carefully against classical heuristics.

Future work should investigate improved QUBO scaling, alternative penalty formulations, post-processing of decoded samples, constrained quadratic models, and richer greenhouse models with additional control variables such as ventilation, shading, humidity, and $\mathrm{CO}_2$. The present benchmark uses a coarse sampling interval of $\Delta t=1$~h, which is sufficient for a structured proof-of-concept benchmark but not intended to represent the full temporal resolution required in operational greenhouse climate control. Practical controllers would require shorter control intervals to capture intra-hour dynamics and actuator constraints.

\begin{appendices}
    
\section{Full greenhouse toy-model specification}
\label{app:toy-model-details}

This appendix provides the full specification of the greenhouse-inspired toy model underlying the benchmark in Section~\ref{sec:greenhouse-toy-model}. The benchmark is intentionally simplified, but it retains a two-state thermal structure, weather and solar forcing, a thermostat-based baseline policy, and diagnostic growth and energy calculations.

\subsection*{Continuous-time thermal model}
\label{app:toy-continuous-model}

The underlying thermal dynamics are defined in continuous time as
\begin{equation}
\begin{aligned}
\frac{dT_{\mathrm{air}}}{dt}
&=
-(k_{\mathrm{env}}+k_{\mathrm{air,body}})T_{\mathrm{air}}
+
k_{\mathrm{air,body}}T_{\mathrm{body}}\\
&+
k_{\mathrm{env}}T_{\mathrm{out}}
+
K_{\mathrm{heater}}u_{\mathrm{applied}}
+
K_{\mathrm{solar}}S, \\
\frac{dT_{\mathrm{body}}}{dt}
&=
k_{\mathrm{body,air}}T_{\mathrm{air}}
-
k_{\mathrm{body,air}}T_{\mathrm{body}},
\end{aligned}
\label{eq:app_toy_continuous_dynamics}
\end{equation}
where $T_{\mathrm{air}}$ is the greenhouse air temperature, $T_{\mathrm{body}}$ is a slower body or canopy temperature, $T_{\mathrm{out}}$ is the outdoor temperature, $S$ is the solar radiation, and $u_{\mathrm{applied}} \in \{0,1\}$ is the heater action effectively applied to the system. The parameter $k_{\mathrm{env}}$ denotes the heat-exchange coefficient between the greenhouse and the outside environment, $k_{\mathrm{air,body}}$ the coupling coefficient from air to body temperature in the air balance, $k_{\mathrm{body,air}}$ the relaxation coefficient governing how the body temperature follows the air temperature, $K_{\mathrm{heater}}$ the heater gain, and $K_{\mathrm{solar}}$ the solar-gain coefficient.

The continuous-time model is converted to discrete time by zero-order hold, which yields the discrete-time form used in the main paper, cf. \eqref{eq:toy_dynamics}. The benchmark is initialized from
\begin{equation}
\mathbf{x}_0
=
\begin{bmatrix}
T_{\mathrm{air},0}\\
T_{\mathrm{body},0}
\end{bmatrix},
\label{eq:app_toy_initial}
\end{equation}
where $T_{\mathrm{air},0}$ and $T_{\mathrm{body},0}$ denote the initial air and body temperatures.

\subsection*{Exogenous input profiles}
\label{app:toy-inputs}

The exogenous inputs are outdoor temperature, solar radiation, and electricity price. Solar radiation is prescribed by a daytime half-sine profile,
\begin{equation}
S_k=
\begin{cases}
S_{\max}\sin\!\left(
\pi\frac{h_k-h_{\mathrm{sunrise}}}{h_{\mathrm{sunset}}-h_{\mathrm{sunrise}}}
\right),
&
h_{\mathrm{sunrise}}\le h_k \le h_{\mathrm{sunset}},
\\[1ex]
0,
&
\text{otherwise},
\end{cases}
\label{eq:app_toy_solar}
\end{equation}
where $S_{\max}$ is the peak solar radiation, $h_k$ is the hour of day corresponding to step $k$, $h_{\mathrm{sunrise}}$ is the sunrise hour, and $h_{\mathrm{sunset}}$ is the sunset hour.

The outdoor temperature profile is generated as a smooth day--night forcing signal. Let
\[
h_k = (h_{\mathrm{start}} + k\Delta t)\bmod 24
\]
denote the hour of day. In the benchmark, $h_{\mathrm{start}}=6$, $T_{\mathrm{out,night}}=2^\circ\mathrm{C}$, $T_{\mathrm{out,day}}=8^\circ\mathrm{C}$, $t_{\mathrm{out,min}}=0$, and $t_{\mathrm{out,max}}=12.5$. The outdoor temperature is defined by
\begin{equation}
T_{\mathrm{out},k}
=
\begin{cases}
T_{\mathrm{out,night}}
+
\frac{T_{\mathrm{out,day}}-T_{\mathrm{out,night}}}{2}
\left[
1-\cos\!\left(
\pi \frac{h_k-t_{\mathrm{out,min}}}{t_{\mathrm{out,max}}-t_{\mathrm{out,min}}}
\right)
\right],
&
t_{\mathrm{out,min}}\le h_k\le t_{\mathrm{out,max}},
\\[2ex]
T_{\mathrm{out,night}}
+
\frac{T_{\mathrm{out,day}}-T_{\mathrm{out,night}}}{2}
\left[
1+\cos(\pi \xi_k)
\right],
&
\text{otherwise},
\end{cases}
\label{eq:app_toy_tout}
\end{equation}
where
\begin{equation}
\xi_k=
\begin{cases}
\dfrac{h_k-t_{\mathrm{out,max}}}{24-(t_{\mathrm{out,max}}-t_{\mathrm{out,min}})},
& h_k>t_{\mathrm{out,max}},
\\[2ex]
\dfrac{h_k+24-t_{\mathrm{out,max}}}{24-(t_{\mathrm{out,max}}-t_{\mathrm{out,min}})},
& h_k<t_{\mathrm{out,min}}.
\end{cases}
\label{eq:app_toy_tout_xi}
\end{equation}

The electricity tariff follows a weekday three-zone structure,
\begin{equation}
p_k=
\begin{cases}
0.08, & 0\le h_k<8,\\
0.12, & 8\le h_k<10,\\
0.18, & 10\le h_k<14,\\
0.12, & 14\le h_k<18,\\
0.18, & 18\le h_k<22,\\
0.12, & 22\le h_k<24.
\end{cases}
\label{eq:app_toy_tariff}
\end{equation}

\subsection*{Thermostat baseline}
\label{app:toy-thermostat}

To generate a baseline control trajectory for comparison and illustration, the simulator uses a thermostat with hysteresis. Let $T_{\min}$ denote the lower comfort bound and $d_{\mathrm{band}}$ the deadband width. The commanded heater action is
\begin{equation}
u_k=
\begin{cases}
1, & T_{\mathrm{air},k}<T_{\min},\\
0, & T_{\mathrm{air},k}>T_{\min}+d_{\mathrm{band}},\\
u_{k-1}, & \text{otherwise},
\end{cases}
\label{eq:app_toy_thermostat}
\end{equation}
where the third case preserves the previous heater command inside the hysteresis band. The model may also include an optional lag between the commanded heater action $u_k$ and the applied heater action $u_{\mathrm{applied},k}$; in the no-lag benchmark setting one sets
\begin{equation}
u_{\mathrm{applied},k}=u_k.
\label{eq:app_toy_applied_nolag}
\end{equation}

This thermostat is used only to generate reference trajectories in the simulator; it is not the optimization method studied in the main benchmark.

\subsection*{Growth and energy diagnostics}
\label{app:toy-diagnostics}

The benchmark uses the reduced growth proxy defined in the main text by \eqref{eq:toy_growth_temp}--\eqref{eq:toy_growth}. For completeness, the interval growth contribution is
\begin{equation}
G_k = g_k \Delta t,
\label{eq:app_toy_interval_growth}
\end{equation}
and cumulative growth is obtained by summing $G_k$ over the horizon.

Likewise, the interval energy use and cost are given by \eqref{eq:toy_energy_cost}, and cumulative energy use and cost are obtained by summing the interval quantities over the horizon.

The number of heater switching events is computed as
\begin{equation}
N_{\mathrm{switch}}
=
\sum_{k=1}^{H-1}|u_k-u_{k-1}|,
\label{eq:app_toy_switches}
\end{equation}
and feasibility is assessed with respect to the air-temperature bounds in \eqref{eq:toy_bounds}.

\subsection*{Benchmark instance and overview figure}
\label{app:toy-benchmark-instance}

\begin{figure}[h]
    \centering    
    \includegraphics[width=0.92\textwidth]{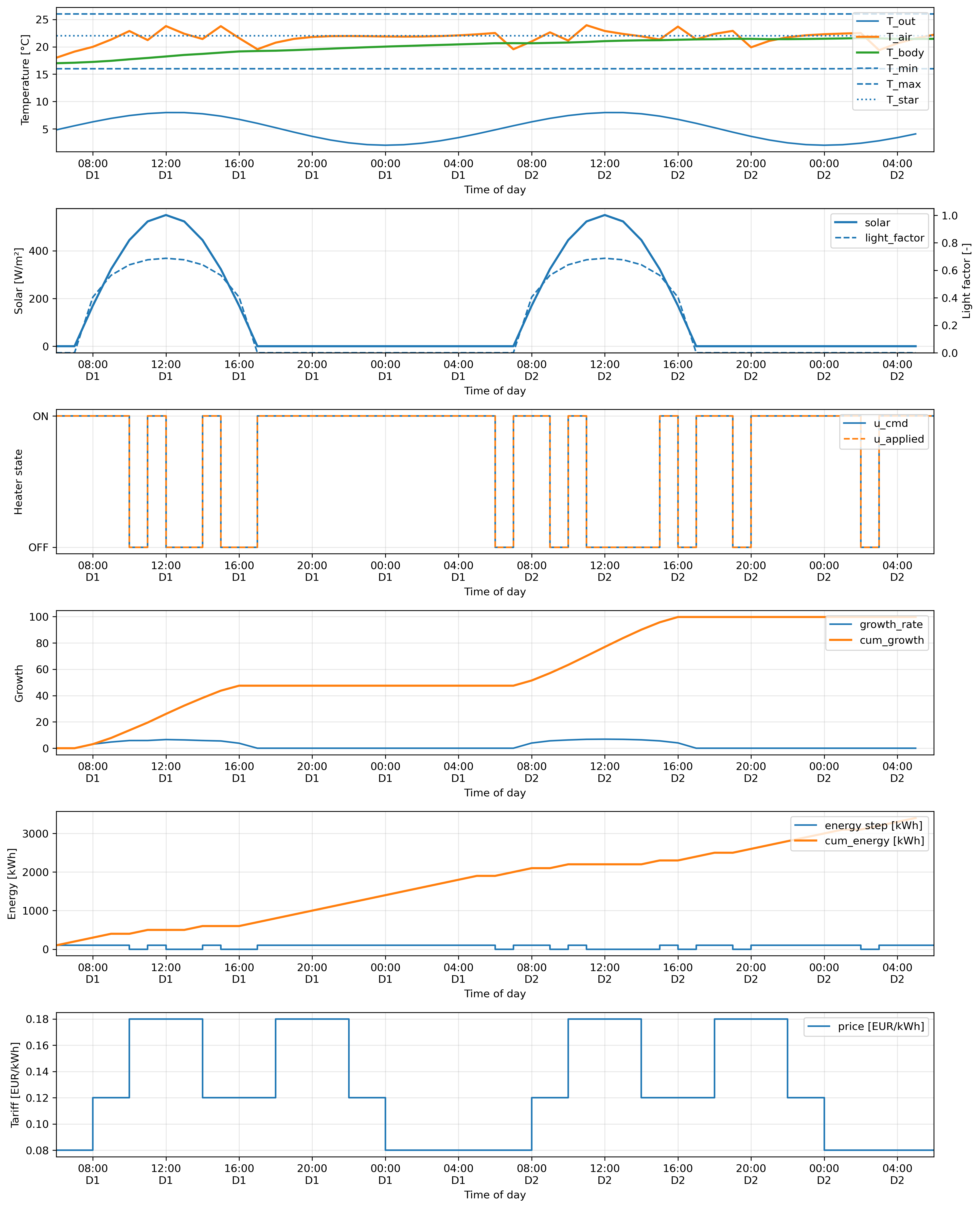}
    \caption{Overview of the greenhouse toy-model simulator over the benchmark horizon, with the horizontal axis expressed as clock time. From top to bottom, the panels show: (i) the thermal trajectories $T_{\mathrm{out}}$, $T_{\mathrm{air}}$, and $T_{\mathrm{body}}$ together with the bounds $T_{\min}$, $T_{\max}$, and the reference temperature $T^\star$; (ii) the prescribed solar radiation and the derived light factor; (iii) the heater control variable $u_k \in \{0,1\}$, where $u_k=0$ denotes the heater-off state and $u_k=1$ denotes the heater-on state, shown for both the commanded and applied heater signals; (iv) the instantaneous growth proxy and cumulative growth; (v) the interval energy use and cumulative energy consumption; and (vi) the electricity price tariff used to weight energy consumption over the day.
    In the plotted simulator overview instance, the horizon is $H=48$ with $\Delta t=1$~h, the initial conditions are $T_{\mathrm{air},0}=18.0^\circ\mathrm{C}$ and $T_{\mathrm{body},0}=17.0^\circ\mathrm{C}$, the thermal parameters are $k_{\mathrm{env}}=0.18~\mathrm{h}^{-1}$, $k_{\mathrm{air,body}}=0.25~\mathrm{h}^{-1}$, $k_{\mathrm{body,air}}=0.06~\mathrm{h}^{-1}$, $K_{\mathrm{heater}}=4.0~^\circ\mathrm{C}/\mathrm{h}$, and $K_{\mathrm{solar}}=0.0045~^\circ\mathrm{C}/(\mathrm{h}\,\mathrm{W}\,\mathrm{m}^{-2})$; the thermostat uses $T_{\min}=16.0^\circ\mathrm{C}$ and $d_{\mathrm{band}}=1.0^\circ\mathrm{C}$; the solar profile uses $S_{\max}=550.0~\mathrm{W}\,\mathrm{m}^{-2}$, $h_{\mathrm{sunrise}}=7.0$, and $h_{\mathrm{sunset}}=17.0$; the growth model uses $\alpha=0.4$, $K_{\mathrm{light}}=250.0~\mathrm{W}\,\mathrm{m}^{-2}$, $g_{\max}=10.0$, $\eta=0.18$, and $T^\star=22.0^\circ\mathrm{C}$; and the energy calculation uses $P_{\mathrm{heater}}=100.0$~kW with $\lambda_E=1.5$ and a time-dependent tariff $p_k$.}
    \label{fig:toy_model_overview}
\end{figure}

Figure~\ref{fig:toy_model_overview} illustrates a representative simulator instance and the main diagnostic outputs of the toy model. This figure is used to visualize the model dynamics and is not the main $H=24$ solver benchmark instance.

\section{Additional trajectory diagnostics for the $H=24$ benchmark}
\label{app:additional_trajectory_diagnostics}

\begin{figure}[t]
    \centering
    \includegraphics[width=\textwidth]{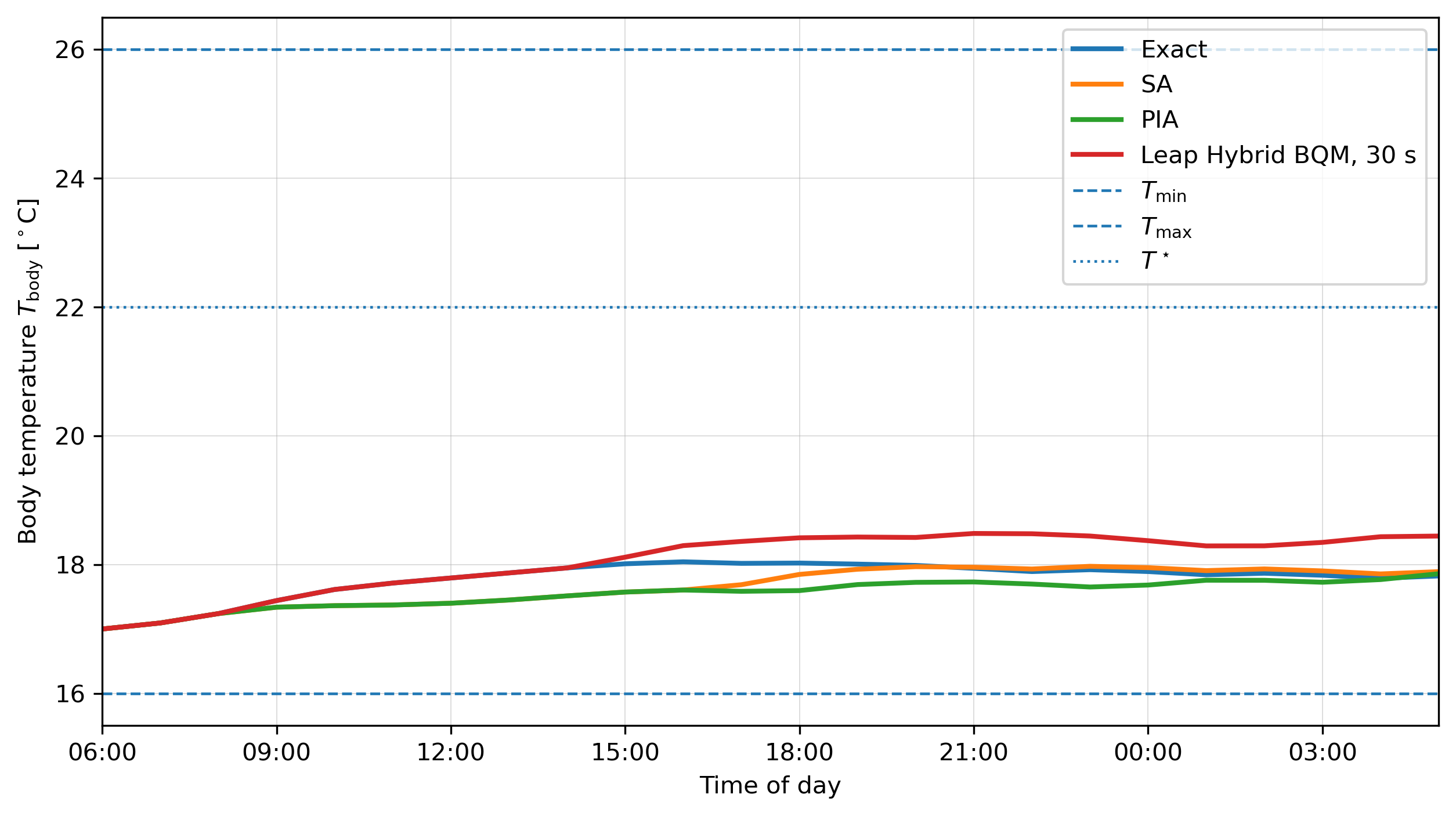}
    \caption{Body-temperature trajectories for the best feasible decoded solutions of the $H=24$ benchmark. The body temperature is smoother than the air temperature and mainly serves as a secondary diagnostic of the two-state thermal model.}
    \label{fig:app_body_temperature_comparison}
\end{figure}

Figure~\ref{fig:app_body_temperature_comparison} shows the body-temperature trajectories corresponding to the best feasible decoded solutions of the $H=24$ benchmark. Compared with the air temperature, the body temperature evolves more smoothly because of the slower thermal dynamics of the second state. The figure is therefore included as an additional diagnostic rather than as a main result.

\begin{figure}[t]
    \centering
    \includegraphics[width=\textwidth]{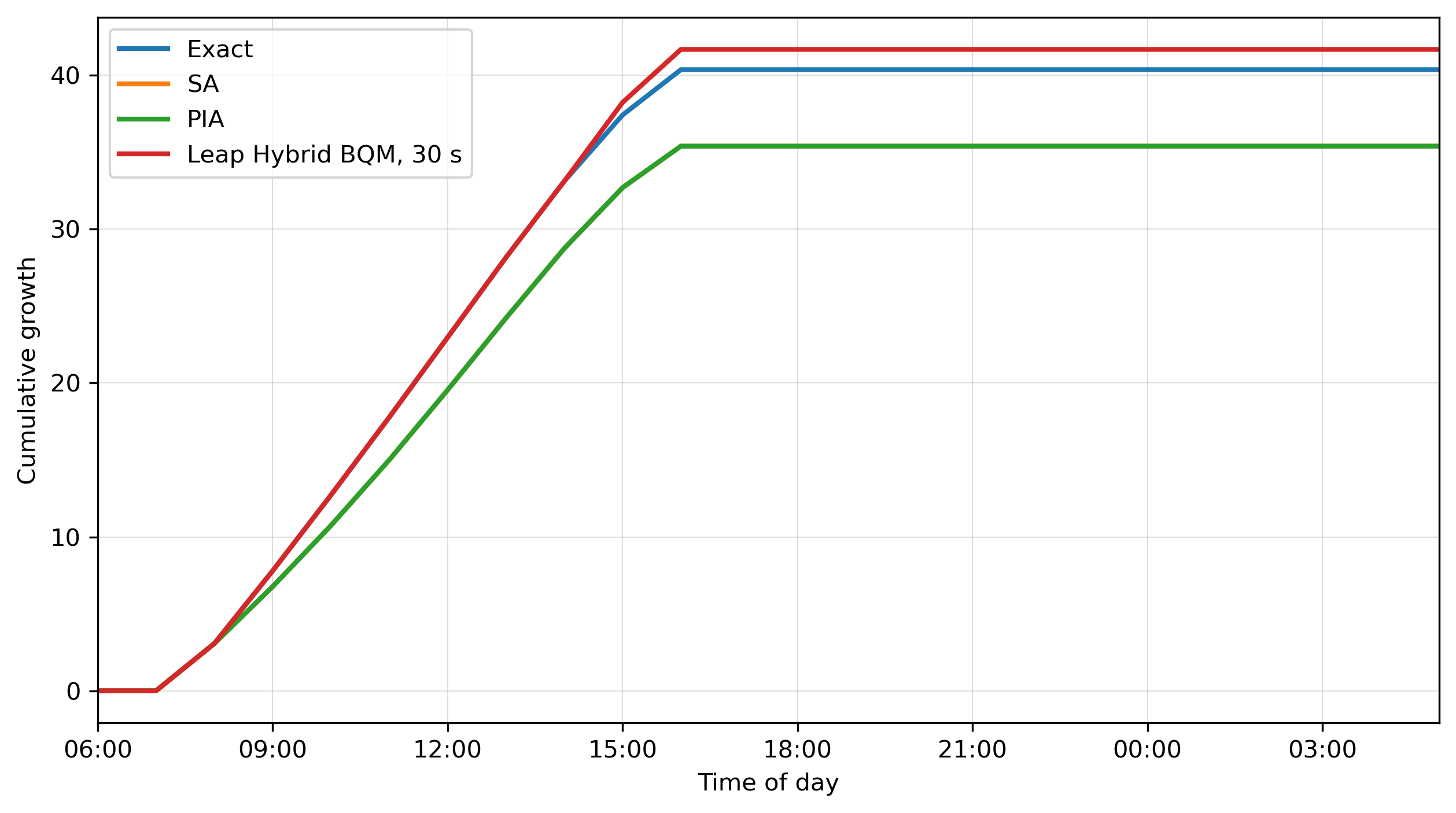}
    \caption{Cumulative growth proxy for the best feasible decoded solutions of the $H=24$ benchmark. The figure is provided as an additional diagnostic because the main growth--energy--switching trade-off is reported quantitatively in Table~\ref{tab:main_kpi_comparison}.}
    \label{fig:app_growth_comparison}
\end{figure}

Figure~\ref{fig:app_growth_comparison} shows the cumulative growth proxy for the best feasible decoded solutions. The hybrid solution obtains a comparatively high final growth value, but this is achieved with increased energy consumption. This supports the interpretation that the relevant benchmark criterion is the full decoded objective, not growth alone.

\end{appendices}

\backmatter

\bmhead{Acknowledgements}

This project was funded by the German Federal Ministry for Economic Affairs
and Energy (BMWE) under grant no. 03EGTTH024 -- QUSMOS.

\section*{Declarations}

\bmhead{Funding}

This work was supported by the German Federal Ministry for Economic Affairs
and Energy (BMWE) under grant no. 03EGTTH024 -- QUSMOS.

\bmhead{Competing interests}

The authors are affiliated with QUSMOS, which is developing quantum-inspired and quantum-ready optimization solutions for industrial applications. Beyond this affiliation, the authors declare no competing interests directly related to the results reported in this manuscript.

\bmhead{Author contributions}

H.A. conceived the study, developed the greenhouse-inspired benchmark formulation, implemented the QUBO model and solver workflows, performed the numerical analysis, and wrote the initial manuscript draft. M.B.Z. contributed to the model development, interpretation of results, and manuscript revision. Both authors reviewed and approved the final manuscript.

\bmhead{Data availability}

The benchmark parameter files and result data used to generate the reported tables and figures will be provided as supplementary material during review and are available from the corresponding author upon reasonable request after publication.

\bmhead{Code availability}

The code used for the simulations, QUBO construction, solver execution, and benchmark analysis will be provided as supplementary material during review and is available from the corresponding author upon reasonable request after publication.

\bibliographystyle{sn-mathphys-num}
\bibliography{greenhouse_refs}

\end{document}